\documentclass[a4paper,11pt]{article}
\usepackage[]{epsfig}
\usepackage{amsfonts}
\usepackage[latin1]{inputenc}
\usepackage[T1]{fontenc}
\usepackage{amsmath}
\usepackage{amssymb}
\usepackage{color}
\usepackage{graphicx}

\newcommand{\be}{\begin{equation}}
\newcommand{\ee}{\end{equation}}
\newcommand{\ba}{\begin{eqnarray}}
\newcommand{\ea}{\end{eqnarray}}

 \newcommand{\bea}{\begin{eqnarray}} \newcommand{\eea}{\end{eqnarray}}

\begin{document}
\title{ Lifshitz fermionic theories  with  $z=2$ anisotropic scaling}
\author{H.~Montani$^a$   and
F.~A.~Schaposnik$^b$\thanks{Associated with CICBA}
\\
~
\\
~
\\
{\normalsize $^a\!$\it Departamento de Ciencias Exactas y Naturales}\\
{\normalsize $\!$\it Unidad Acad\'emica Caleta Olivia}\\
{\normalsize $\!$\it  Universidad Nacional de la Patagonia Austral}\\
{\normalsize $\!$\it 9011 Caleta Olivia, Argentina}\\
{\normalsize $\!$\it }\\
{\normalsize $^b\!$\it Departamento de F\'\i sica, Universidad
Nacional de La Plata}\\ {\normalsize\it Instituto de F\'\i sica La Plata-CONICET}\\
{\normalsize\it C.C. 67, 1900 La Plata,
Argentina}
 }
\date{\hfill}
\maketitle
\begin{abstract}
{We construct fermionic Lagrangians with anisotropic scaling $z=2$, the natural counterpart of the usual
$z=2$  Lifshitz field theories for scalar fields. We analyze the issue of chiral symmetry, construct the Noether axial currents and
discuss the  chiral anomaly giving explicit results for   two-dimensional case. We also exploit
the connection between detailed balance and the dynamics of Lifshitz theories to find different $z=2$ fermionic  Lagrangians and construct their supersymmetric extensions.}
\end{abstract}

\section{Introduction}
Lifshitz field theories exhibit  {an} anisotropic scaling between space and time that
breaks Lorentz symmetry.
 Introduced in the context of condensed matter phenomena \cite{H}-\cite{Frad}, they
are  {also}  actively investigated in connection with  problems in gravity \cite{Horava0}-\cite{Horava2} {and in} particle physics (see \cite{Alex} and reference therein).

 The anisotropic gauge scaling for spatial coordinates $x$ and time $t$ takes the form
\ba
x &\to&  bx, \nonumber\\
t &\to& b^zt
\label{scale}\ea
where $b\in \mathbb{R}$ is the scaling factor and $z\in \mathbb{Z}$ the degree of anisotropy   which in
condensed matter systems is {identified}    as a dynamical critical exponent. The  $z=1$ case corresponds to the
ordinary scaling law in the conformal group. In connection with this, it becomes natural to {study,} within
the context of gauge/gravity correspondence,
metrics with anisotropic scaling which could be dual to Lifshitz field theories  in the same way as
AdS metrics allow to establish duality leading to  strongly coupled conformal field theories \cite{Kachru}.

A prototype of a Lifshitz field theory is a $(d+1)$-dimensional  scalar theory with dynamics governed by the action
\be
S = \frac12\int d^dxdt \left( \left({\partial_t\phi}\right)^2 -  \kappa^2 \phi\left(-\nabla^2\right)^z\phi
\right)
\label{action}
\ee
Here $\nabla^2 \equiv \partial_i\partial_i$ is the spatial Laplacian. The scaling dimensions are
$[x^i] = -1$ and $[t]= -z$, with $[\phi] = (d-z)/2$ and $[\kappa] = z-2$.

Models with action (\ref{action}), particularly in the   $z=2$ case interest  condensed matter physicists
 because they  exhibit a line of fixed points parametrized by $\kappa$  and reproduce
 the phase diagrams of known materials \cite{Gri}. Other $z\ne1$ bosonic field theories have been  {discussed,}
 and also their gravity duals have been investigated using the holographic principle with a
 bulk geometry given (in Poincar\'e coordinates) by the metric
\be
  ds^2 = L^2\left(-\frac{dt^2}{u^{2z}} + \frac{d\vec x^2}{u^2} + \frac{du^2}{u^2}\right)
  \label{metric2}
  \ee
   which for $z=1$   becomes the usual AdS metric.

Defining fermionic Lifshitz theories is an open issue. Already in  refs.\cite{AF,Frad}, dealing with   conformal quantum critical points,  the problem is addressed  for the case of   $z=2$ scaling  by defining
a ground state wave functional for the Hamiltonian of a $2+1$ system in terms of the $(1+1)$   Euclidean Dirac action, but the discussion
in this work is restricted to a bosonized version of the theory.
Self-interacting fermionic Hamiltonians leading to $z=2$ dynamical critical exponent have been constructed in \cite{fk} in $1+1$ dimensions following an approach different to  that based in theories defined as in (\ref{action}).
 More recently fermion theories with $z=3$ have been studied in connection with its renormalizability, chiral anomaly and
renormalization group flow \cite{Anselmi1}-\cite{AlexandreU}.

 In this work we undertake the construction of $z=2$ fermionic Lifshitz theory following a different way.  Having in mind that the Dirac equation can be seen as the ``square root'' of the Klein-Gordon one, we firstly start from   the bosonic action (\ref{action})  and derive a {$\gamma$}-matrices  algebra which naturally leads to a  Lifshitz fermion model in $d+1$ space-time dimensions.
In doing so, we introduce the square root of the Laplacian as the source of the anisotropic
scaling. We consider the z = 2 case which corresponds to $(-\nabla^2)^{1/2}$  but the power $z/2$ could be equally considered giving rise to theories with  arbitrary  even $z$. We shall also discuss the issue of  chiral symmetry  for massless fermions and consider the case in which they couple to an Abelian gauge field. We work out the corresponding classical Noether currents and we also  compute the chiral anomaly in  $1+1$ space-time dimensions which, as it is well-known,
 is closely related to  bosonization, one of the issues of relevance in condensed matter applications of Lifshitz theories  \cite{AF}.

 Secondly, we  present an approach based on the connection  between the principle of detailed balance   and the dynamics of Lifshitz theories stressed in refs. \cite{Horava0}-\cite{Horava2}. The main ingredient in this approach is  stochastic field theory quantization provides, a useful tool in the construction Lifshitz field theory actions. Indeed,
 starting from the relativistic action for scalar fields  {in $d$ Euclidean dimensions} stochastic quantization  leads
  to a $d+1$ effective action with the fictitious time derivatives  appearing quadratically  and those of  the $d$ remaining derivatives quartically. Following this approach, we shall analyze the stochastic quantization of fermionic theories and construct various Dirac-Lifshitz actions with different anisotropy scaling $z$ including $z=2$.

 The paper is organized as follows.  We establish in Section 2 a connection between  $z=2$ bosonic
 and fermionic theories  leading to a consistent $z=2$ Dirac-Lifshitz action in $d+1$ space-time dimensions. In section 3 we
 discuss the issue of chiral symmetry finding the classically conserved axial currents
an calculating the quantum  chiral anomaly in the particular case of a 1+1 dimensional space. Working within the stochastic quantization approach,
 we discuss in Section 4 the construction of  alternative $z=2$ fermionic theories in $d+1$ dimensions as well as their supersymmetric extension.  {Finally we present  in Section 5 a discussion of our results.}

\section{Connecting bosonic and fermionic $z=2$ Lifshitz theories}

Consider a Lifshitz bosonic action of the form (\ref{action}) in $d+1$
Euclidean spacetime   with a $z=2$ scaling
\begin{equation}
S_{L}= \int d^{d+1}x\left( \frac{1}{2}\left( \partial _{0}\varphi
\right) ^{2}+\frac{\kappa ^{2}}{2}\left( \nabla ^{2}\varphi \right)
^{2}\right)  \label{actionm}
\end{equation}%
or
\begin{equation}
S=\frac{1}{2}\int d^{d+1}x\left( \frac{1}{2}\left( \partial _{0}\varphi
\right) ^{2}-\frac{\kappa ^{2}}{2}\left( \partial _{i}\varphi \right) \nabla
^{2}\left( \partial _{i}\varphi \right) \right) \; , \;\;\;\; i=1,2,\ldots, d
\label{disa}
\end{equation}%
It will be useful for our purposes to write action (\ref{disa}) in the form
\begin{equation}
S=\frac{1}{2}\int d^{d}x~\partial _{\mu }\varphi ~G_{\mu \nu }\left( \nabla
^{2}\right) \partial _{\nu }\varphi  \label{cra}
\end{equation}%
with  $G_{\mu \nu }$ defined as
\begin{equation}
(G_{\mu \nu })=\left(
\begin{array}{cc}
I_{1\times 1} & 0_{1\times n} \\
0_{n\times 1} & -\kappa ^{2}\nabla ^{2}I_{n\times n}%
\end{array}%
\right)
\end{equation}%
The integral on the r.h.s. of eq.(\ref{cra}) can be regarded as a degenerate scalar product on the space of vector
fields on $M$,
\begin{equation*}
(\;,)_{M}:\mathfrak{X}\left( M\right) \otimes \mathfrak{X}\left( M\right)
\longrightarrow \mathbb{R}
\end{equation*}%
such that
\begin{equation}
(V(x),W(x))_{M}=\int d^{d}x\left( V_{0}W_{0}-\kappa ^{2}V_{i}\nabla
^{2}W_{i}\right)
\end{equation}
The associated group of {isometries} can be obtained from the invariance
relation%
\begin{equation*}
\left( \Lambda \left( V\right) ,\Lambda \left( W\right) \right) _{M}=\left(
V,W\right) _{M}
\end{equation*}%
which explicitly reads%
\begin{equation}
\Lambda _{0\mu }\Lambda _{0\nu }-\kappa ^{2}\Lambda _{i\mu }\Lambda _{i\nu
}\nabla ^{2}=G_{\mu \nu }\left( \nabla ^{2}\right)   \label{relations 0}
\end{equation}%
{Taking $\mu ,\nu =0$, we get}%
\begin{equation*}
\begin{array}{ccc}
\Lambda _{i0}\Lambda _{i0}=0 & , & \Lambda _{00}^{2}=1%
\end{array}%
\end{equation*}%
which implies{\   }%
\begin{equation*}
\begin{array}{ccc}
\Lambda _{i0}=0 & , & \Lambda _{00}=\pm 1%
\end{array}%
\end{equation*}%
{respectively. Alternatively, by taking $\mu =i$ and }$\nu ${$=j$ in equation (\ref{relations 0}), we determine
the relations defining the remaining components, namely,%
\begin{equation*}
\begin{array}{ccc}
\Lambda _{0i}=0 &  & \Lambda _{ki}\Lambda _{kj}=\delta _{ij}%
\end{array}%
\end{equation*}%
implying that $\Lambda _{ij}$ are the entries of an orthogonal matrix. In
summary, the final form of $\Lambda $ is
\begin{equation}
\begin{array}{ccccc}
\Lambda _{00}=1 & , & \Lambda _{0i}=\Lambda _{i0}=0 & , & \Lambda
_{ki}\Lambda _{kj}=\delta _{ij}%
\end{array}
\label{isometry}
\end{equation}%
meaning that that isometries of $G_{\mu \nu }\left( \nabla ^{2}\right) $
reduce to rotations in the spatial sector and time reversal map.

One can extend this framework to the case of fermionic theories in a natural way
by defining
\textquotedblleft local\textquotedblright\ gamma matrices $\Gamma $
satisfying
\begin{equation*}
\{\Gamma _{\mu },\Gamma _{\nu }\}=2G_{\mu \nu }
\end{equation*}%
An appropriate choice for these gamma matrices is
\begin{equation*}
\begin{array}{ccc}
\Gamma _{i}=\kappa (-\nabla ^{2})^{1/2}\gamma _{i} & , & \Gamma _{0}=\gamma
_{0}%
\end{array}%
\,, \;\;\; i=1,2,\ldots d
\end{equation*}%
where $(\gamma _{a})\equiv (\gamma _{0},\gamma _{i})$ satisfy the ordinary
Euclidean-space Clifford algebra
\begin{equation*}
\{\gamma _{a},\gamma _{b}\}=2\delta _{ab}  \,, \;\;\; a=0, 1, \ldots d
\end{equation*}%
and $(-\nabla ^{2})^{1/2}$ is the nonlocal operator which can be defined as in
\cite{Caffa}   (see also references   \cite{Seeley}-\cite{Bollini}).
\begin{equation}
(-\nabla ^{2})^{1/2}f\left( x\right) =P.V.\int_{\mathbb{R}^{d}}\frac{f\left(
x\right) -f\left( y\right) }{\left\vert x-y\right\vert ^{d+1}}d^{d}y
\end{equation}%
Defined in this way the square root of the Laplacian is an elliptic pseudodifferential operator of order
$1$.
  However, a better insight of the non local operator $(-\nabla ^{2})^{1/2}$ is attained in the context of the harmonic extensions of a compactly supported function on  $\mathbb{R}^{n}$ to  $\mathbb{R}^{n}\times
\left( 0, \infty \right)$, as in ref. \cite{Caffa}, where it appears as a map from the Dirichlet-type data problem to the Neumann-type
data one. Moreover, it can be shown that it is self-adjoint and positive-definite (see \cite{Shubin}-\cite{Tan} and references therein).  In Appendix   we give a brief description of this approach and some properties of this operator.
From now on, we shall use a more compact notation for the square root of the Laplacian naming it by $T$
\be
T:=(-\nabla^2 )^{1\!/2}
\label{T operator}
\ee
For further developments, it is useful  to work out the action of the operator  $T$ on imaginary exponentials. From the identity
\be
- \nabla^2 \exp(ikx) = |k|^2 \exp(ikx)
\label{explained}
\ee
and  the natural definition of the square root Laplacian it can be established that
\be
T \exp(ikx)  =  |k| \exp(ikx)
\label{valorabs}
\ee
(see the Appendix for a proof of this result).
Let us point that, strictly speaking, since the   operator symbol $|k|$  is not smooth at { the origin, the problem} should be handled inserting an appropriate cutoff function.

We can now propose a vierbein-like connection between $\Gamma $ and $\gamma $
\begin{equation}
\left( E_{\mu }^{~a}\right) =\left(
\begin{array}{cc}
I & 0\cr0 & \kappa TI%
\end{array}%
\right) \delta (x-y)\,,\;\;\left( E_{a}^{~\mu }\right) =\left(
\begin{array}{cc}
I & 0\cr0 & -\frac{1}{\kappa }TI%
\end{array}%
\right) \delta (x-y)
\end{equation}%
 lowering and raising  indices with   $G_{\mu \nu }$ and $\delta _{ab}$ so that
\begin{equation*}
\left( E_{\mu }^{~a}E_{a\nu }\right) =\left( G_{\mu \nu }\right)
\;,\;\;\;\left( E_{a}^{~\nu }E_{\nu b}\right) =(\delta _{ab})
\end{equation*}%
with products  including an integration over space-time.

This set of matrices $\left\{ \Gamma _{\mu }\right\} $ naturally induces
a definition of  a {Dirac-Lifshitz  equation} in the form
\begin{equation}
(i\Gamma _{\mu }\partial _{\mu }-m)\psi =0  \label{Dcrazy}
\end{equation}%
Here $m$ has dimensions of mass squared.
In terms of the ordinary $\gamma _{\mu }$ matrices eq.(\ref{Dcrazy}) reeds
\begin{equation}
\left( i\gamma _{0}\partial _{0}+i\kappa T(\gamma
_{i}\partial _{i})-m\right) \psi =0
\label{DLDL}
\end{equation}%
This is nothing but the Euler-Lagrange equation for action
\begin{equation}
S_{DL}=\int d^{d+1}x\bar{\psi}\left( i\gamma _{0}\partial _{0}+i\kappa T\gamma _{i}\partial _{i}-m\right) \psi   \label{DL}
\end{equation}%
These Dirac-Lifshitz fields shares the anisotropic gauge scaling $\left( \ref%
{scale}\right) $ for $z=2$. From this expressions we define the {%
Dirac-Lifshitz operator}%
\begin{equation}
\gamma _{\mu }D_{\mu }=i\gamma _{0}\partial _{0}+i\kappa T\gamma _{i}\partial _{i}-m
\label{LDop}
\end{equation}

It
is interesting to note that, as in the ordinary Dirac equation case,  one can proceed to ``square'' the Dirac Lifshitz equation looking for a resulting Klein-Gordon-like equation. Indeed,  multiplying (\ref{Dcrazy}) equation by the
operator $(i\Gamma _{\mu }\partial _{\mu }+m)$ and using the identity (see Appendix)
\begin{equation}
T\partial _{i}=\partial
_{i}T   \label{42012}
\end{equation} we get the usual $z=2$ Lifshitz equation  for scalar fields,  subject of many recent investigations
\begin{equation}
\left( \partial _{0}^{2}-\kappa ^{2}\left( \nabla ^{2}\right)
^{2}-m^{2}\right) \psi =0
\label{17}
\end{equation}

This is precisely the equation of motion arising from the $z=2$ bosonic action (\ref{actionm})
and in this respect Lagrangian (\ref{DL})  seems to be a natural candidate to study   Lifshitz fermionic  models in connection with the problems of the conformal quantum critical points
\cite{AF}.

\section{Chiral symmetry and the $z=2$ Dirac-Lifshitz Lagrangian}
 {The issue of chiral symmetry and the quantum anomaly of $z=3$ Lifshitz theories with massless fermions coupled to a gauge field in $3+1$ space-time dimensions has been thoroughly studied in \cite{Dhar:2009dx},  \cite{Bakas:2011nq}-\cite{Bakas}. Using the path-integral framework the resulting anomaly ${\cal A}$ was shown to be identical to the usual relativistic calculation and  this is likely to be related to its topological character in the sense
that ${\cal A}$ is metric indepedent and its integral -the topological charge- does not depend on local details but only on global properties. The calculation for the $z=2$ Lifshitz defined in the precedent section  seems more involved since
it implies handling the square root of the Dirac operator both in the action  and as the natural regulator in a sensible definition of the path-integral measure. We shall now discuss this issue.
\subsection{The free massless fermion theory}
 In the $m=0$ case the free fermion Lifshitz action (\ref{DL})
 is invariant under chiral rotations
 \be
\psi \to \exp(\gamma_5 \epsilon ) \psi \; , \;\;\;\;\;
\bar \psi \to \bar  \psi\exp(  \gamma_5 \epsilon )
\label{promoting}
\ee
 There should then be, at the classical level, a  continuity equation for the chiral current and an associated  conserved chiral charge.  To obtain their explicit form one can proceed as follows. One promotes the global change to a local (infinitesimal for simplicity) one
 \bea
\psi &\to& \psi' = \exp(\epsilon(x) \gamma_5) \psi \approx (1 + \epsilon(x)\gamma_5)\psi
\nonumber\\
\bar \psi &\to& \bar \psi' = \bar  \psi\exp(\epsilon(x) \gamma_5) \approx \bar \psi (1 + \epsilon(x)\gamma_5)
\label{promoting2}
\eea
under which the action should necessarily change,
\be
\delta S_{DL} =
\int   d^{d+1}x  \bar  \psi(1 + \gamma_5\epsilon(x))\left( i\gamma^{0}\partial _{0}+i\kappa T\gamma ^{i}\partial _{i}-m\right) (1 + \gamma_5\epsilon(x)) \psi
- S_{DL}\nonumber
\ee
After some work one finds
\bea
\delta S_{DL} &=& \int d^{d+1}x  i\bar\psi \gamma^1\gamma_5\psi \partial_0\epsilon + \int d^{d+1}x  i\kappa \bar\psi T\gamma^i\gamma_5
\psi \partial_i\epsilon
\nonumber\\
 &=&
 - i\int d^{d+1}x \bar\psi \left(\gamma^0\gamma_5\psi \partial_0 +\kappa T \gamma ^{i}\gamma_5  \right) \psi \epsilon
 \label{dDL}
\eea
For constant $\epsilon$ this term should be absent and then we get a continuity equation and the explicit
form of the chiral current
\begin{align}
  \partial_\mu j^\mu_5 &= 0\nonumber\\
  j^0_5 = \bar\psi \gamma^0 \gamma_5\psi\, ,\hspace{1 cm} & \hspace{1 cm}
\!j^i_5 = \kappa\bar\psi  \gamma^i \gamma_5T\psi
\end{align}
As it was to be expected,  the chiral charge coincides with the usual (Dirac theory) one,
\be
Q_5 = \int d^dx \bar\psi \gamma^0 \gamma_5\psi = \int d^dx  \psi^\dagger  \gamma_5\psi
\ee

There is of course a $U(1)$ global invariance with associated current
\be
 j^0 = \bar\psi \gamma^0  \psi\, ,\hspace{1 cm} \hspace{1 cm}
\!j^i = \kappa\bar\psi  \gamma^i  T\psi
\label{analog}
\ee
so that again, the conserved $U(1)$ charge coincides with the usual one,
\be
Q = \int d^dx \bar\psi \gamma^0 \psi = \int d^dx  \psi^\dagger \psi
\ee
\subsection*{Coupling fermions to $U(1)$ gauge fields}

In order to couple massless $z=2$ Lifshitz  fermions  to a $U(1)$
gauge field background $A_{\mu }$, we introduce  the
  extension of the
Lifshitz-Dirac operator (\ref{LDop})
\begin{equation}
\mathcal{D}=\gamma ^{\mu }\mathcal{D}_{\mu }[A]=\gamma _{0}D_{0}\left[ A%
\right] +\kappa \gamma _{1}D_{1}\left[ A\right] \mathcal{T}\left[ A\right]
\label{LDform}
\end{equation}%
where we have defined
\begin{equation}
\mathcal{T}\left[ A\right] =(D_{j}\left[ A\right] D_{j}\left[ A\right]
)^{1/2}
\end{equation}%
%
with covariant derivatives   given by
\begin{equation}
D_{0}=i\partial _{0}+eA_{0}\;,\;\;\;D_{i}=i\partial _{i}+eA_{i}
\label{deri2}
\end{equation}%
Some properties of operator ${\cal T}[A]$ are discussed in the Appendix. In
particular, $\mathcal{T}[A]$ is an Hermitian operator so that $\mathcal{D}$
defined as in (\ref{LDform}) is also Hermitian.

A $z=2$ Dirac-Lifshitz action can be compactly written in terms of the
Dirac-Lifshitz operator $\mathcal{D}_{\mu }[A]$ as
\begin{equation}
S_{DL}[A]=\int d^{d+1}x\,\bar{\psi}\gamma ^{\mu }\mathcal{D}_{\mu }[A]\psi
\label{simple}
\end{equation}%
As in the free case, this action remains invariant under global chiral
transformations (as well as under local $U(1)$ gauge transformation).
So, there should be a  classically  conserved axial current.
Concerning the quantum level, one should expect an anomaly which can
shall calculate within the path-integral framework following the well-honored
  Fujikawa's approach
\cite{Fujikawa}-\cite{Fujikawa2}. We start from the partition
function written in the form
\begin{equation}
Z=\int D\bar{\psi}D\psi \exp (-S[A])
\end{equation}%
and consider the infinitesimal local path-integral change of variables
given by eq.(\ref{promoting2}). 
The action in the exponential now gets, as in the case of the free fermion
classical action, a contribution proportional to $\partial _{\mu }\epsilon $%
; and concerning the path-integral measure, there is a Fujikawa Jacobian $%
J_{5}[\epsilon ]$ . So, the partition function changes to
\begin{equation}
Z=\int D\bar{\psi}^{\prime }D\psi ^{\prime }J_{5}[\epsilon ]\exp
\left(\vphantom{\frac12}\!\!-S[A]-\delta S[A;\partial _{\mu }\epsilon ]\right)
\end{equation}%
As in the free fermion  case, we shall  show below that $\delta S$ can be
accommodated as
\begin{equation}
\delta S[A;\partial _{\mu }\epsilon ]=\int d^{d}x(\partial _{\mu }\epsilon
(x))j_{5}^{\mu }  \label{probar}
\end{equation}%
so that $Z$ can be written as
\begin{equation}
Z=\int D\bar{\psi}^{\prime }D\psi ^{\prime }J_{5}[\epsilon ]\exp
\left(\vphantom{\frac12}\!\!-S[A]-\epsilon (x)(\partial _{\mu }j_{5}^{\mu })\right)
\end{equation}%
Since the generating functional cannot depend on the parameter $\epsilon
(x) $ introduced by the change of variables one has
\begin{equation}
\frac{1}{Z}\frac{\delta Z}{\delta \epsilon (x)}=0
\end{equation}%
which leads to the anomaly equation
\begin{equation}
\langle \partial _{\mu }j_{5}^{\mu }\rangle =\mathcal{A}  \label{ecuacionA}
\end{equation}%
where
\begin{equation}
\mathcal{A}=\frac{\delta \log J_{5}}{\delta \epsilon (x)}
\end{equation}%
Let us now prove eq.(\ref{probar}). The action $S[A]$ can   be written as
\begin{equation}
S[A]=\int d^{d+1}x\bar{\psi}\left( \gamma _{0}D_{0}+\kappa \gamma ^{i}\left(
D_{i}[A]\mathcal{T}[A]\right) \right) \psi
\end{equation}%
After the change of variables (\ref{promoting2}) one   has
\begin{equation}
\delta S[A]=S_{0}[A]+S_{1}[A] -S[A]
\end{equation}%
with
\begin{eqnarray}
S_{0}[A] &=&\int d^{d+1}x\bar{\psi}(1+\gamma _{5}\epsilon (x))\gamma
^{0}D_{0}(1+\gamma _{5}\epsilon (x))\psi  \notag \\
S_{1}[A] &=&\kappa \int d^{d+1}x\bar{\psi}(1+\gamma _{5}\epsilon (x))\gamma
^{i}D_{i}\mathcal{T}[A](1+\gamma _{5}\epsilon (x))\psi
\end{eqnarray}%
The term $S_{0}$ contains the same variation arising in an ordinary Dirac theory and
can then be written in the form
\begin{equation}
S_{0}=-\int d^{d+1}x\epsilon (x)\partial _{i}\left( \bar{\psi}\gamma
^{0}\gamma _{5}\psi \right) +\int d^{d}x\bar{\psi}i\gamma ^{0}D_{0}\psi
\end{equation}

In order to handle $\delta S_{1}$ in a similar way, we first integrate by
parts
\begin{eqnarray}
\delta S_{2}[A]&=&\kappa \int d^{d+1}xD_{i}^{\ast }\bar{\psi}(1+\gamma
_{5}\epsilon (x))\gamma ^{i}\mathcal{T}[A](1+\gamma _{5}\epsilon (x))\psi
\nonumber\\
&&-\kappa \int d^{d+1}x\bar{\psi}\gamma ^{i}D_{i}\mathcal{T}[A]\psi
\end{eqnarray}%
finally getting
\be
\delta S_{2}=-i\kappa \int d^{d+1}x(\partial _{i}\epsilon )\bar{\psi}\gamma
^{5}\gamma ^{1}\mathcal{T}[A]\psi =i\kappa \int d^{d+1}x\epsilon (x)\partial
_{i}\left( \bar{\psi}\gamma ^{5}\gamma ^{1}\mathcal{T}[A]\psi \right)
\end{equation}%
Putting all this together we have for the complete variation of the action
\begin{equation}
\delta S=\int d^{d+1}x\epsilon (x)\partial _{\mu }j_{5}^{\mu }
\end{equation}%
with
\begin{equation}
j_{5}^{0}=\bar{\psi}\gamma ^{0}\psi \,,\hspace{1cm}\hspace{1cm}%
\!j_{5}^{i}=\kappa \bar{\psi}\gamma ^{i}\mathcal{T}[A]\psi  \label{analogous}
\end{equation}%
which differs from the free fermionic case because $j_{5}^{i}$ contains the
operator $\mathcal{T}[A]={(D_{i}D_{i})^{1/2}}$ instead of $T={(-\nabla
^{2})^{1/2}}$ and hence depends on $A_{i}$.

We have now to compute the Fujikawa Jacobian associated to the chiral change
of variables. For simplicity we shall consider the $1+1$ dimensional case in which
Lagrangian in action (\ref{simple}) takes the simple form
\begin{equation}
L=\bar{\psi}\gamma ^{0}(D_{0})\psi +\kappa\bar{\psi}\left( \gamma
^{1}D_{1}\mathcal{T}[A]\right) \psi
\end{equation}%
Our choice of  gamma matrices is
\begin{equation}
\gamma _{0}=\left(
\begin{array}{cc}
0 & 1 \\
1 & 0%
\end{array}%
\right) \;,\;\;\;\;\gamma _{1}=\left(
\begin{array}{cc}
0 & i \\
-i & 0%
\end{array}%
\right) \;,\;\;\;\;\gamma _{5}=i\gamma _{0}\gamma _{1}=\left(
\begin{array}{cc}
1 & 0 \\
0 & -1%
\end{array}%
\right)
\end{equation}

Following  Fujikawa's approach, we consider a change of variables as in (\ref{promoting2}) and  calculate the associated Jacobian $J^{5}[\delta
\epsilon (x)]$ which gives
\begin{equation}
J_{5}[\delta \epsilon (x)]=\exp \left( 2\int d^{2}x\sum_{n}\varphi
_{n}^{\dagger }(x)\gamma _{5}\varphi _{n}(x)\delta \epsilon (x)\right)
\end{equation}%
From the fact that the generating functional cannot depend on the parameter $%
\epsilon (x)$ introduced by the change of variables
\begin{equation}
\frac{\delta Z}{\delta \epsilon (x)}=0
\end{equation}%
one then gets
\begin{equation}
\partial _{\mu }j_{5}^{\mu }=\frac{\delta \log J_{5}}{\delta \epsilon (x)}=%
\mathcal{A}  \label{ecuacionA}
\end{equation}%
with $\mathcal{A}$ given by the ill-defined formula
\begin{equation}
\mathcal{A}=2\sum_{n}\varphi _{n}^{\dagger }(x)\gamma _{5}\varphi
_{n}(x)=\lim_{y\rightarrow x}2\,\mathrm{tr}\,\gamma _{5}\delta ^{(2)}(x-y)
\label{16-A}
\end{equation}%
which thus requires an appropriate regularization
\begin{equation}
\mathcal{A}_{reg}=\left. \vphantom{\frac12}\lim_{y\rightarrow x}2\mathrm{tr}%
\gamma _{5}\delta ^{(2)}(x-y)\right\vert _{reg}  \label{Alinda}
\end{equation}

A practical regularization approach is the so call heat-kernel method which
consist in introducing an appropriate regulating operator $R^{2}$ and a
cut-off mass $M$ in the form
\begin{equation}
\mathcal{A}_{reg}=\lim_{y\rightarrow x}2\mathrm{tr}\left( \gamma _{5}\exp
(-R^{2}/\Lambda \lbrack M]^{2})\delta ^{(2)}(x-y)\right)  \label{56}
\end{equation}%
where $\Lambda \lbrack M]=M^{q}$ with $q$ chosen so that $[\Lambda ]=[R]=q$.
A finite result for the anomaly is obtained by interchanging trace and limit
\begin{equation}
\mathcal{A}=\lim_{M\rightarrow \infty }\lim_{y\rightarrow x}2\mathrm{tr}%
\left( \gamma _{5}\exp (-R^{2}/\Lambda ^{2})\delta ^{(2)}(x-y)\right)
\label{formulita}
\end{equation}%
Now, the regulator should be chosen so as to preserve the basic invariances
of the theory, in the present case gauge invariance and also the invariance
under the anisotropic scaling (\ref{scale}). This makes natural to use as
regulator the operator appearing in the Lagrangian so that $R^{2}={\mathcal{D%
}}{\mathcal{D}}$ and then one has to choose $\Lambda \lbrack M]=M^{4}$.

In two space-time dimensions the Dirac-Lifshitz operator reduces to
\begin{equation}
\mathcal{D}=i\gamma _{0}D_{0}+{\kappa \gamma _{1}D_{1}[A]}\mathcal{T}{[A]}
\end{equation}%
and ${\cal T}[A]$ is simply
\begin{equation}
\mathcal{T}[A]=(D_{1}[A]D_{1}[A])^{1/2}
\end{equation}%
 In the Appendix we show that ${\cal T}[A]$ can be also written in the form
\begin{equation}
\mathcal{T}\left[ A\right] =W^{-1}\left[ A_1\right] TW\left[ A_1%
\right]%
\label{D1 - 0}
\end{equation}
where
\be
W[A_1] = \exp \left( -ie\int^{x}A_{1}\left(
y\right) dy\right)
\ee
At this point, it is important to note that one can gauge away one component of the gauge potential  just by a gauge transformation.  {In particular,  we shall consider} $W[A_1]$ as a
 $U(1)$-gauge group element $W\left[ A_{1}\right] $  so that
\begin{equation*}
\left( A_{0},A_{1}\right) \longrightarrow \left( W\left[ A_{1}\right]
A_{0}W^{-1}\left[ A_{1}\right] ,0\right)
\end{equation*}%
which is formally implemented as%
\begin{equation}
A_{\mu }\longrightarrow A_{\mu }+W\left[ A_{1}\right] \left( i\partial _{\mu
}\right) W^{-1}\left[ A_{1}\right]
\label{2012}
\end{equation}%
This gauge will greatly simplify calculations leading to a result that can be trivially extended to
a general gauge.

The regulator $\mathcal{D}^{2}$ can be now written in the form
\begin{equation}
\mathcal{D}^{2}=ID_{0}D_{0}-i\kappa \gamma _{5}\left[ D_{0},W^{-1}\left[
A_{1}\right] \left( i\partial _{1}\right) TW\left[ A_{1}\right] \right]
+\kappa ^{2}IW^{-1}\left[ A_{1}\right] \left( -\partial _{1}^{2}\right) ^{2}W%
\left[ A_{1}\right]
\end{equation}%
Under  the gauge transformation (\ref{2012}) $D_{\mu }$ changes according to
\begin{equation*}
D_{\mu }\longrightarrow W\left[ A_{1}\right] D_{\mu }W^{-1}\left[ A_{1}%
\right]
\end{equation*}%
so that the regularizing operator becomes %
\begin{eqnarray}
W\left[ A_{1}\right] \mathcal{D}^{2}W^{-1}\left[ A_{1}\right]  &=&IW\left[
A_{1}\right] D_{0}D_{0}W^{-1}\left[ A_{1}\right] +\kappa ^{2}I\left(
-\partial _{1}^{2}\right) ^{2}  \notag \\
&&-i\kappa \gamma _{5}\left[ W\left[ A_{1}\right] D_{0}W^{-1}\left[ A_{1}%
\right] ,\left( i\partial _{1}\right) T\right]
\end{eqnarray}
{with this and the  $\delta ^{(2)}(x-y)$  function written in term of the plane wave basis
eq.(\ref{formulita}) becomes
\begin{equation}
\mathcal{A}_{reg}=\frac{1}{\left( 2\pi \right) ^{2}}\lim_{M\rightarrow
\infty }\lim_{y\rightarrow x}\int d^{2}p~2\mathrm{tr}\left( \gamma
_{5}e^{-ip_{\mu }y_{\mu }}e^{-\left( W\left[ A_{1}\right] \mathcal{D}%
^{2}W^{-1}\left[ A_{1}\right] \right) /M^{4}}e^{ip_{\mu }x_{\mu }}\right)
\end{equation}%
Taking now the limit }$y\rightarrow x$, {the operator content of }$\mathcal{A}%
_{reg}$ becomes
\begin{equation*}
e^{-ip_{\mu }x_{\mu }}W\left[ A_{1}\right] \mathcal{D}^{2}W^{-1}\left[ A_{1}%
\right] e^{ip_{\mu }x_{\mu }}=I\Phi \left( A,p\right) -i\kappa \gamma
_{5}\Theta \left( A,p\right) +p_{0}^{2}+\kappa ^{2}Ip_{1}^{4}
\end{equation*}%
Here   we have  used that%
\begin{equation*}
e^{-ip_{1}x_{1}}W\left[ A_{1}\right] =W\left[ A_{1}+\frac{p_{1}}{e}\right]
\end{equation*}
and introduced $\Phi$ and $\Theta$ defined as
\begin{eqnarray*}
\Phi \left( A,p\right)  &=&W\left[ A_{1}+\frac{p_{1}}{e}\right] \left(
D_{0}^{2}-2p_{0}\left( i\partial _{0}\right) \right) W^{-1}\left[ A_{1}+%
\frac{p_{1}}{e}\right]  \\
&& \\
\Theta \left( A,p\right)  &=&\left[ W\left[ A_{1}+\frac{p_{1}}{e}\right]
\left( i\partial _{0}\right) W^{-1}\left[ A_{1}+\frac{p_{1}}{e}\right]
,e^{-ip_{1}x_{1}}\left( i\partial _{1}\right) Te^{ip_{1}x_{1}}\right]
\end{eqnarray*}

Making the substitution $p_{0}=M^{2}k_{0}$ and $p_{1}=Mk_{1},$the anomaly
becomes{%
\begin{eqnarray}
\mathcal{A}_{reg} \!\!&=& \!\!\frac{1}{\left( 2\pi \right) ^{2}}\lim_{M\rightarrow
\infty}\int d^{2}p~2M^{3}
 \left(\exp-\left( k_{0}^{2}+\kappa ^{2}k_{1}^{4}\right)\right) \nonumber\\
\!\!&& \!\!\times
 \mathrm{tr}\left( \gamma _{5}\exp\left(-\left( I\Phi \left(
A,M^{2}k_{0},Mk_{1}\right)
 -i\kappa \gamma _{5}\Theta \left(
A,M^{2}k_{0},Mk_{1}\right) \right) /M^{4}\right)\right) \nonumber\\
\end{eqnarray}%
By $M$ power counting
in the argument of the exponential,  one concludes that only
 the linear term in  $\Theta \left( A,M^{2}k_{0},Mk_{1}\right) $ survives after taking $M\rightarrow \infty $ and
the $\gamma$-matrix trace.
In order to carry out the explicit calculation of  this contribution, recall that  }
 before differentiating with respect to $\epsilon(x)$ the regulated Jacobian
includes an integral over 2-dimensional space and hence one can use the
integration by parts properties of $T$ (see the Appendix) and then differentiate, thus obtaining%
\begin{equation*}
\Theta \left( A,M^{2}k_{0},Mk_{1}\right) =ieM\left\vert k_{1}\right\vert
F_{01}
\end{equation*}%
 {The anomaly can  then be finally written in a gauge invariant way as {%
\begin{equation}
\mathcal{A}_{reg}=-\frac{e}{4\pi }\varepsilon ^{\mu \nu }F_{\mu \nu }
\end{equation}%
which coincides with the result for the ordinary relativistic two dimensional
\begin{equation}
\partial _{\mu }j_{5}^{\mu }=-\frac{e}{4\pi }\varepsilon ^{\mu \nu }F_{\mu
\nu }
\label{anomacero}
\end{equation}%
}As observed in refs. \cite{Dhar:2009dx} and \cite{Bakas:2011nq} for $z=3$ models
 in $3+1$ dimensions (and then generalized to arbitrary values of $z$ in \cite%
{Bakas}), the coincidence between the anomaly of  relativistic $z=1$  and
Lifshitz $z\neq 1$ theories is not surprising in retrospect since the anomaly $%
\mathcal{A}$ is related to a topological density that coincides with a total
divergence, so that  eq.(\ref{anomacero}) is the only possible result for the
anomaly, including the universal coefficient, independently of the value of $z$.
 To confirm this en the $1+1$ dimensional case,  we briefly describe the $1+1$ anomaly calculation for the $z=3$ model, with the action in the form
usually taken in the literature,
\be
S = \int d^dx dt \bar \psi \gamma^\mu {\cal D}^{(3)}_\mu  \psi
\ee
Here the Dirac-Lifshitz operator is given by
\be
 {{\cal D}^{(3)}=}\gamma^\mu {\cal D}^{(3)}_\mu = \gamma^0D_0 + \frac12 \gamma^i\left( D_i (-D_k D_k)  + (-D_kD_k)  D_i
\right)
\label{LDform3}
\ee
with the covariant derivatives defined as in (\ref{deri2}).
Again, at the classical the there is a conserved Noether charge associated to a global chiral
transformation. At the quantum level, one should take into account the change in the fermionic measure
and one ends with the anomaly equation (\ref{ecuacionA})
\be
{{\cal A} = \lim_{M\to \infty} \lim_{y\to x} 2
{\rm tr}\left(\gamma_5 \exp(-{{\cal D}^{(3)}} {{\cal D}^{(3)}}/M^6)\delta^{(2)}(x-y)\right)}
\label{formulita2}
\ee

As before, the regulator has been chosen so as to preserve the basic invariances of the
theory (i.e. gauge invariance and also the anisotropic scaling invariance) and hence $\Lambda[M]$ in (\ref{56}) should be
$\Lambda[M] = M^6$.

In  $1+1$ dimensions the Dirac-Lifshitz operator reduces to
\begin{equation*}
{\mathcal{D}}^{(3)}=i\gamma _{0}D_{0}-\kappa \gamma _{1}D_{1}^{3}
\end{equation*}
and ${\mathcal{D}}^{(3)2}$ is given by %
\begin{equation}
{\mathcal{D}}^{(3)2}=ID_{0}^{2}-i\kappa \gamma _{5}\left[ D_{0},D_{1}^{3}\right]
+\kappa ^{2}ID_{1}^{6}  \label{D cuadrado bakas}
\end{equation}%
We shall not repeat the details of the calculation that follow closely those that leading to
(\ref{anomacero}) in the $z = 2$ case. We just quote the only non-vanishing $M \to \infty$ term
in ${\cal A}$
which instead of integral (\ref{anomacero}) is now given by
\begin{equation*}
\mathcal{A}=-3e\kappa \frac{2}{\left( 2\pi \right) ^{2}}F_{01}\left( \int
d^{2}k~~\mathrm{e}^{-\left( k_{0}^{2}+\kappa ^{2}k_{1}^{6}\right)
}~k_{1}^{2}\right)
\end{equation*}%
Wich can be easily integrated to give
\begin{eqnarray*}
\mathcal{A} &=&-3e\kappa \frac{2}{\left( 2\pi \right) ^{2}}\sqrt{\pi }F_{01}\left( \int
k_1^2 dk_1~\mathrm{e}^{-\kappa ^{2}k_{1}^{6}}~\right)= -\frac{e}{2\pi} F_{01} =
 -\frac{e}{4\pi}\varepsilon^{\mu\nu} F_{\mu\nu}
\end{eqnarray*}%
which again coincides  with the ordinary result for the two-dimensional anomaly,

\section{Lifshitz fermions and  stochastic quantization}

It has been recenlty observed that field theory models with the anisotropic scaling (\ref{scale})  can be thought of as stemming from a stochastic quantization process \cite{Horava}-\cite{Horava2}, \cite{Dijk}-\cite{Orl}. In this way, a $d+1$ dimensional  bosonic Lifshitz theory like that defined by action (\ref{actionm}), with $z=2$ can be thought off  as descending from a free bosonic action in $d$ Euclidean dimensions,
\be
S^{(d)} = \int d^dx \partial_i\phi \partial_i\phi \; , \;\;\;\; i=1,2\ldots,d
\ee
{ The dynamics in extra time variable $\tau$ is ruled by a Langevin equation, namely a first order diferential equation with a dissipative (drift) term and driven by a source of stochastic noise. As far as the quantization of a $d$-dimensional theory is concerned (the asymptotic equilibrium state), there exist some arbitrariness in the election of this stochastic process. One may take some profit from this freedom to build up theories in $d+1$ dimensions with different behaviours under "time" and space scaling.

In the case of the bosonic field (\ref{actionm}), to get a $z=2$ Lifshitz theory one introduce an extra dependence of the fields on this extra time variable, $\phi(x) \to \phi(x,\tau)$, and compose the Langevin equation with $\delta S^{(d)}/\delta \phi$ playing the role of a drift force, and a white noise source as the stochastic force\cite{Parisi}. }
 The  case of fermionic fields can be treated analogously \cite{Fukai}-\cite{DT}, with $\psi$ and $\bar\psi$ taken as independent fields, each one with its own
Langevin equation
\ba
\frac{\partial \psi(x,\tau)}{\partial \tau} &=& -  \frac{\delta S[\bar \psi,\psi]}{\delta\bar\psi} + \eta(x,\tau) \nonumber\\
\frac{\partial\bar \psi(x,\tau)}{\partial \tau} &=&   \frac{\delta S[\ \psi,\psi]}{\delta \psi} + \bar\eta(x,\tau)
\label{sto}
\ea
where $\eta,\bar\eta$ are anticommuting fermionic gaussian noises satisfying
\be
\langle \eta_\alpha(x,\tau)\rangle =  \langle \bar\eta_\alpha(x,\tau)\rangle = 0
\ee
\be
\langle\eta_\alpha(x_1,\tau_1)\bar\eta_\beta (x_2,\tau_2)\rangle = 2\delta(t_1-t_2)\delta^{(d)} (x_1-x_2)
\delta_{\alpha\beta}
\label{delta}
\ee
with $\alpha,\beta$ the spinor indices. Such normalization arises from the following definition of
the (Fokker-Planck) generating functional
\ba
Z[J,\bar J] \!\!&=&\!\! \int D\bar\eta D\eta \exp\left(
-\frac12\int dx \int_0^\tau d\tau' \bar\eta \eta
\right) \exp\left(
\int dx \int_0^\tau d\tau'(\bar\eta J + \bar J \eta)\right)
\nonumber\\
&=& \exp\left(2\int dx \int_0^\tau d\tau' \bar J J\right)
\label{FP}
\ea
where spinor indexes have been omitted and $J,\bar J$ are fermionic external sources.

Now, as already noted in \cite{Fukai}-\cite{DT}, when $S$ is taken as the Dirac action
one should in general
include an appropriate kernel $K(x,y)$ to ensure that the relaxation process is such that the system tends to the equilibrium
when $\tau \to \infty$.  {Different kernels produce stochastic process which can differ in the behaviour under scaling. In any case,}  Langevin equations (\ref{sto}) are then replaced by
\ba
\frac{\partial \psi(x,\tau)}{\partial \tau} &=& -  \int dy K(x,y)\frac{\delta S [\bar \psi,\psi]}{\delta\bar\psi(y,\tau)} + \eta(x,\tau) \nonumber\\
\frac{\partial\bar \psi(x,\tau)}{\partial \tau} &=&  \int dy   \frac{\delta S [\ \psi,\psi]}{\delta \psi(y,\tau)} K(y,x)+ \bar\eta(x,\tau)
\label{stoq}
\ea
and eq.\ (\ref{delta}) by
\be
\langle\eta_\alpha(x_1,\tau_1)\bar\eta_\beta (x_2,\tau_2)\rangle = 2\delta(t_1-t_2)K(x_1-x_2)
\delta_{\alpha\beta}
\label{delta1}
\ee
  {The usual choice for $K$
\be
K(x,y) = (i\not\!\partial_x + m)\delta(x-y)
\ee
in the case $S$ is the Dirac action for a massive fermion,
\be
S = \int d^dx \bar\psi(x)( i\not\!\partial - m)\psi(x)
\ee
 gives rise to the Langevin equations
\ba
\frac{\partial \psi(x,\tau)}{\partial \tau} &=& (\nabla^2 - m^2)\psi + \eta(x,\tau) \nonumber\\
\frac{\partial\bar \psi(x,\tau)}{\partial \tau} &=&   (\nabla^2 - m^2)\bar\psi + \bar\eta(x,\tau)
\label{stoK}
\ea
where a $z=2$ anisotropic scaling turns evident.}

 {Let us now come back to the general fermionic stochastic process defined by the Langevin equations (\ref{delta1}).} In the presence of the kernel $K$ the partition function associated to (\ref{FP})  takes the form
\be
Z = \int\!\! D\bar\eta D\eta \exp\!\left(\!-\frac12 \int\!d^dx d^dy  d\tau\bar\eta(x,\tau) K^{-1}(x,y)\eta(y,\tau)\!
\label{Zf}
\right)
\ee
We now proceed to  change   variables from $\bar\eta, \eta$  to $\bar \psi, \psi$
\be
Z = \int\!\! D\bar\psi D\psi {\det}^{-1}\frac{\delta \bar\eta}{\delta\bar\psi}
{\det}^{-1}\frac{\delta \eta}{\delta\psi}
\exp\!\left(\!-\frac12 \int\!d^dx d^dy  d\tau\bar\eta(x,\tau) K^{-1}(x,y)\eta(y,\tau)\!
\right)
\ee
Using Langevin equations (\ref{stoK}) written in the form
\ba
K^{-1}\frac{\partial \psi}{\partial \tau} &=& -  \frac{\delta S }{\delta\bar\psi} + K^{-1} \eta \nonumber\\
\frac{\partial\bar \psi }{\partial \tau}K^{-1}  &=&   \frac{\delta S }{\delta \psi} + \bar\eta K^{-1}
\label{stocompact}
\ea
we end with
\ba
 Z &=&\!\! \int\!\! D\bar\psi D\psi  {\det}^{-1}  \left(
 K^{-1}\frac{\partial  }{\partial \tau} -  \frac{\delta^2 S[\bar \psi,\psi]}{\delta\bar\psi\delta\psi}
 \right) {\det}^{-1}
 \left(
K^{-1} \frac{\partial}{\partial \tau} +  \frac{\delta^2 S[\bar \psi,\psi]}{\delta\psi\delta\bar\psi}
 \right)\nonumber\\
 &\times & \!\!\exp \!\!\left(-\frac12\int d^dxd\tau\!\! \left( \frac{\partial\bar \psi}{\partial \tau}K^{-1} -
  \frac{\delta S[\ \psi,\psi]}{\delta \psi}
 \right)\!\!\left(
 \frac{\partial \psi(x,\tau)}{\partial \tau} + K \frac{\delta S[\bar \psi,\psi]}{\delta\bar\psi}
 \right)\right) \nonumber\\
\label{ZZZ}
\ea
where we have used the identity
\be
\frac{\delta}{\delta \psi_\beta} K^{-1}_{\alpha\gamma}\partial_\tau\psi_\gamma = K^{-1}_{\alpha,\beta}\partial_\tau
\ee
Determinants in (\ref{ZZZ}) can defined as the (regularized) product of eigenvalues of the
equation
\be
\left(\left(K^{-1}\right)^{\alpha\beta} \frac{\partial}{\partial\tau} - \frac{\delta^2S}{\delta\bar\psi^\alpha\delta\psi^\beta}\right)\phi_i^\beta = \lambda \phi_i^\alpha
\ee
and a similar one for the other factor. These determinants can then be
 represented as   quadratic path-integrals over  two  complex
commuting fields $\phi_i^\alpha\, (i=1,2)$ carrying a spinorial index
 which can be identified with  the spinorial ghosts for ghosts \cite{vanN}.
With this,  the partition function $Z$  can be written as
\be
Z = \int D\bar\psi D\psi D\bar \phi_1 D\bar\phi_2 D\phi_1 D\phi_2 \exp\left(-\int dx d\tau L_{eff}[\bar\psi, \psi, \phi_1,\phi_2]
\right)
\label{15}
\ee
where
\ba
L_{eff} &=& \frac12 \left( \frac{\partial\bar \psi(x,\tau)}{\partial \tau}K^{-1} -
    \frac{\delta S[\ \psi,\psi]}{\delta \psi}
 \right) \left(
 \frac{\partial \psi(x,\tau)}{\partial \tau} +  K\frac{\delta S[\bar \psi,\psi]}{\delta\bar\psi}
 \right)
\nonumber\\
 && + \bar\phi_1 \left(
K^{-1} \frac{\partial  }{\partial \tau} -  \frac{\delta^2 S[\bar \psi,\psi]}{\delta\bar\psi\delta\psi}
 \right)\phi_1  + \bar\phi_2  \left(
K^{-1} \frac{\partial}{\partial \tau} +  \frac{\delta^2 S[\bar \psi,\psi]}{\delta\psi\delta\bar\psi}
 \right)\phi_2\nonumber\\
 \label{16}
 \ea
 or
 \ba
L_{eff}
 &=& \frac12 \frac{\partial\bar \psi(x,\tau)}{\partial \tau} K^{-1}\frac{\partial \psi(x,\tau)}{\partial \tau}
-   \frac12\frac{\delta S[\bar \psi,\psi]}{\delta \psi} K \frac{\delta S[\bar \psi,\psi]}{\delta\bar\psi} \nonumber\\
&& +\bar \phi_1 \left(
 K^{-1}\frac{\partial  }{\partial \tau} -  \frac{\delta^2 S[\bar \psi,\psi]}{\delta\bar\psi\delta\psi}
 \right)\phi_1  +  \bar\phi_2  \left(
K^{-1} \frac{\partial}{\partial \tau} +  \frac{\delta^2 S[\bar \psi,\psi]}{\delta\psi\delta\bar\psi}
 \right)\phi_2\nonumber\\
  \label{166}
\ea
We can see now how the choice of the stochastic process,  {defined by a kernel  K, gives rise} to different scaling
behaviours. Indeed, starting with a $d$ -dimensional free massive fermion Dirac action  and using a trivial kernel
\be
K(x,y) = \delta(x-y)
\ee
 (this being consistent for $m \ne 0$
the fermionic part of the effective fermion Lagrangian (\ref{166}) will have  second derivatives both in time and space variables
\be
L_{fer} = \frac12 \frac{\partial\bar \psi(x,\tau)}{\partial \tau} \frac{\partial \psi(x,\tau)}{\partial \tau}
+ \bar\psi \left(\nabla^2 + m^2)\right)\psi
\ee
which corresponds to a Lifshitz theory wit $z=1$, i.e a theory with isotropic scaling.
 If one instead chooses as $d$ dimensional action a   Klein-Gordon fermionic action
 \be
 S = \int d^dx \bar\psi(x) (\nabla^2 + m^2)\psi(x)
 \ee
 one  ends with a  $z=2$  Lifshitz theory with Lagrangian
 \be
 L_{fer}^{KG2} =
  \frac12 \frac{\partial\bar \psi(x,\tau)}{\partial \tau} \frac{\partial \psi(x,\tau)}{\partial \tau}
+ \bar\psi \left(\nabla^2 + m^2\right)^2\psi
\label{KG2}
 \ee

We shall end this section by making use of another remarkable attribute of
the stochastic quantization scheme, namely the inherent supersymmetry of the
partition function, in order to supersymmetrize the Lifshitz fermionic models
constructed above \cite{ParSour}.
 Alternatively, and following the ideas introduced in the section 2, we may
consider the $d$ dimensional action for fermions%
\begin{equation*}
S_{LF}=\int d^{d}x\bar{\psi}\left( i\kappa (-\nabla ^{2})^{1/2}\gamma
_{i}\partial _{i}-m\right) \psi
\end{equation*}%
which gives rise through the stochastic quantization approach to the fermionic Lifshitz Lagrangian%
\bea
L_{eff} &=&\frac{1}{2}\frac{\partial \bar{\psi}(x,\tau )}{\partial \tau }\frac{%
\partial \psi (x,\tau )}{\partial \tau }-\frac{1}{2}\bar{\psi}\left( -\kappa
^{2}(-\nabla ^{2})^{1/2}\gamma _{i}\partial _{i}(-\nabla ^{2})^{1/2}\gamma
_{j}\partial _{j}+m^{2}\right) \psi \nonumber\\
&=&
\frac{1}{2}\frac{\partial \bar{\psi}(x,\tau )}{\partial \tau }\frac{%
\partial \psi (x,\tau )}{\partial \tau }+\frac{1}{2}\bar{\psi}\left( \kappa^{2} (\nabla^2)^2
-m^{2}\right) \psi
\eea
which also exhibits a $z=2$ anisotropic behaviour under scaling.


 Starting from partition function  Z as given in (\ref{15}) with Lagrangian (\ref{166}),  we introduce fermionic  auxiliary fields  $\bar F, F$ so that $Z$ becomes \cite{Cha}
\be
Z = \int D\bar\psi D\psi D\phi_1 D\phi_2 D\bar\phi_1 D\bar\phi_2 D\bar F D F \exp\left(-\int d^dx d\tau L_{eff}[\bar\psi, \psi, \phi_1,\phi_2,\bar F, F]
\right)
\label{15n}
\ee
\ba
L_{eff}&=&  2 \bar FK^{-1}F +  i\left( \frac{\partial\bar \psi(x,\tau)}{\partial \tau}K^{-1} -
    \frac{\delta S_f^E[\ \psi,\psi]}{\delta \psi}
 \right) F\nonumber\\
&&  + i \bar F \left(
 \frac{\partial \psi(x,\tau)}{\partial \tau} + K \frac{\delta S_f^E[\bar \psi,\psi]}{\delta\bar\psi}
 \right) +
 \bar\phi_1 \left( K^{-1}\frac{\partial}{\partial \tau} -
    \frac{\delta^2 S_f^E[\bar\psi,\psi]}{\delta\bar\psi\delta \psi}
 \right)\phi_1
\nonumber\\
 &&
+ \bar\phi_2 \left(
K^{-1} \frac{\partial }{\partial \tau} +  \frac{\delta^2 S_f^E[\bar \psi,\psi]}{\delta\psi\delta\bar\psi}
 \right)\phi_2
 \label{16n}
\ea
We now define fermion superfields in the form
\ba
\Psi &=& \psi + \bar\theta \phi_1 + \bar\phi_2\theta +  i\bar\theta\theta F \nonumber\\
\bar \Psi &=& \bar \psi +\bar \theta \phi_2 + \bar \phi_1 \theta +   i\bar\theta\theta \bar F
\ea
and
supercovariant derivatives $D$ and $\bar D$ acting on superfields according to
\be
 D= \frac{\partial}{\partial \bar\theta} -   \theta \frac\partial{\partial\tau} \; , \;\;\;\; \bar D = \frac\partial{\partial \theta}
\ee
\be
D^2 = {\bar D}^2 = 0 \; , \;\;\;\; \{D,\bar D\}= -\frac{\partial}{\partial\tau} =-H
\ee
Supersymmetry generators $Q$ and $\bar Q$ take the form
\be
Q = \frac\partial{\partial\bar \theta}\; , \;\;\;\; \bar Q =  \frac\partial{\partial \theta}
+ \bar\theta\frac\partial{\partial\tau}
\ee
so that
\be
\{Q,\bar Q\} =  \frac\partial{\partial\tau} =  H
\ee
A supersymmetry transformation with parameters $(\epsilon,\bar \epsilon)$ on superfields
reads
\be
{\cal Q}\Psi =  \bar\epsilon Q + \bar Q \epsilon
\ee

In order to write the suppersymmetry transformations in component fields we consider the action of $Q$ and $\bar Q$ on superfields
\ba
Q \Psi &=&
\frac\partial{\partial\bar\theta}
 \left(
  \psi + \bar\theta \phi_1 + \bar\phi_2 \theta+  i\bar\theta\theta F
 \right) = \phi_1 + i\theta F \nonumber\\
 \bar Q \Psi &=& \left( \frac\partial{\partial \theta}
+ \bar\theta\frac\partial{\partial\tau}  \right)
\left(
\psi + \bar\theta \phi_1 + \bar\phi_2 \theta+  i\bar\theta\theta F
\right)\nonumber\\
&=& \bar\phi_2  -\bar \theta F +\bar \theta \dot \psi + \bar \theta \theta \dot {\bar\phi}_2
 \ea
 so that
 \be
 {\cal Q}\Psi = \bar\epsilon\left(  \phi_1 + i\theta F
 \right) + \left(\bar\phi_2  -i\bar \theta F +\bar \theta \dot \psi + \bar \theta \theta \dot {\bar\phi}_2\right)\epsilon
 \ee
 and then
\be
\begin{array}{ll}
\delta \psi = \bar \epsilon \phi_1 +\bar\phi_2   \epsilon  &\hspace{ 2 cm}\delta \phi_1 = \bar \epsilon (\dot \psi - iF)
\\
\delta \phi_2 = i \bar\epsilon F    &\hspace{ 2 cm}\delta F =    -i \dot \phi_2 \epsilon

\end{array}
\ee

Let us consider the action
\be
S_1 = \int d\bar\theta d\theta \bar D\bar\Psi D\Psi
\ee
that written in term of component fields is given by
\ba
S_1 &=& \int  d\bar\theta d\theta
 \frac{\partial}{\partial \theta}
 \left(
\bar \psi +\bar \theta \phi_2 + \bar \phi_1 \theta +   i\bar\theta\theta \bar F
\right)
\nonumber\\&&
\times
\left(\frac{\partial}{\partial \bar\theta} -   \theta \frac\partial{\partial\tau}\right)
\left(\psi + \bar\theta \phi_1 + \bar\phi_2\theta + i \bar\theta\theta F  \right)
\nonumber\\
&=&
\int  d\bar\theta d\theta \left(\bar\phi_1 -i \bar\theta \bar F
\right)
\left(\phi_1 +
i\theta F - \theta\dot\psi +  \bar\theta\theta \dot\phi_1
\right)  \nonumber\\
&=& \bar\phi_1 \dot\phi_1 + \bar F F +i\bar  F \dot{\psi}
\ea

Define also
\be
S_2 = \int d\bar\theta d\theta   \bar D\Psi D\bar\Psi
\ee
or
\ba
S_2 &=& \int d\bar\theta d\theta
\frac{\partial}{\partial \theta} \left(\psi + \bar\theta \phi_1 + \bar\phi_2\theta + i \bar\theta\theta F  \right)
\nonumber\\
&& \times \left(\frac{\partial}{\partial \bar\theta} -   \theta \frac\partial{\partial\tau}\right)
\left(
\bar \psi +\bar \theta \phi_2 + \bar \phi_1 \theta +   i\bar\theta\theta \bar F\right) \nonumber\\
&=& \int d\bar\theta d\theta
\left(
\bar\phi_2 -i\bar\theta F
\right)\left( \phi_2 + i \theta \bar F - \theta\dot{\bar\psi} + \bar\theta\theta \dot\phi_2
\right)\nonumber\\
&=& \bar \phi_2\dot\phi_2 + F\bar F + i F \dot{\bar\psi}
\ea
so that
\be
S_1 + S_2 = \int d\bar\theta d\theta \left(\bar D\bar\Psi D\Psi +
\bar D\Psi D\bar\Psi \right) =  \bar\phi_1 \dot\phi_1 + \bar \phi_2\dot\phi_2 +2 \bar F F + i(\bar F \dot\psi + \dot{\bar\psi}F)
\label{43}
\ee
For the case $K = K^{-1} = I$ the terms in the right hand side exhaust all terms in (\ref{16n}) containing temporal derivatives. For nontrivial $K$ one has just to make the following
change of variables:
\begin{align}
\bar F K^{-1} \to \bar F \; , & \;\;\;\; F \to F
\nonumber\\
\bar\psi K^{-1} \to \bar \psi \; , & \;\;\;\; \psi \to \psi
\nonumber\\
\bar\phi_1 K^{-1} \to \bar \phi_1 \; , & \;\;\;\; \phi_1 \to \phi_1
\nonumber\\
\bar\phi_2 K^{-1} \to \bar \phi_2 \; , & \;\;\;\; \phi_2 \to \phi_2
\end{align}
and the calculation reduces to the $K=K^{-1}$ one.

As for the  terms in (\ref{16n}) depending on action $S$, one has just to add to Lagrangian (\ref{43}) the following term
\be
S_S = \int d\bar\theta d\theta S[\bar\Psi,\Psi]
\ee
Expanding the action in powers of the superfield the only terms that contribute to the $\bar\theta, \theta$
integral are
\be
S_S = \int d\bar\theta d\theta  S[\bar\Psi,\Psi] = \bar F \frac{\delta  S  }{\delta \bar\psi}
+  \frac{\delta  S  }{\delta\psi} F + \bar\phi_1 \frac{\delta^2 S }{\delta\psi\delta\bar\psi} \phi_1
+ {\bar \phi_2}\frac{\delta^2 S }{\delta\psi\delta\bar\psi} \phi_2
\ee
so that the fermionic sector of superaction
\be
{\cal S} = \int d\bar\theta d\theta d^dxd\tau \left(\bar D\bar\Psi D\Psi +
\bar D\Psi D\bar\Psi  +S_f^E[\bar\Psi,\Psi]\right)
\ee
coincides with the action defined from Lagrangian (\ref{166}).

\newpage

\section{Discussion}
  {The source of the anisotropic scaling in the Lagrangian formulation of Lifshitz theories is the choice of  spatial
and time derivatives of different orders. This is generally achieved by
replacing first order spatial derivatives in   relativistic Lagrangians by the Laplacian operator (or integer powers of it)  while leaving unchanged the term with time-derivatives. Concerning bosonic theories, whose standard relativistic Lagrangian is quadratic in spatial derivatives,  such replacement   gives rise to a $z=2$ anisotropy (or $z>2$ for higher Laplacian powers \cite{Alex}).
{For} fermionic theories, application
of  the same strategy to
the Dirac Lagrangian leads to an anisotropy degree $z=3$ (or $z > 3$ \cite{Bakas}). Now,  Lifshitz theories have recently received
much attention in connection with condensed matter systems for which the $z=2$ case is of particular interest. In particular, in \cite{AF},\cite{Frad} the  issue of a $z=2$ fermionic systems in $2+1$ dimension  has been
related to a bosonic $z=2$ Hamiltonian, invoking a  bosonization approach.}

 {We have proposed in this work another route to study Lifshitz fermion
actions with $z=2$ in $d + 1$ space-time dimensions, by introducing the square root of the Laplacian, and its natural extension when fermions are coupled to gauge fields. Roughly, it consists in replacing spatial derivatives $\partial _{i}$
by the operator {$(-\nabla ^{2})^{1/2}$}$\partial _{i}$, which naturally
leads both to bosonic and fermionic $z=2$ Lifshitz theories.}
We have stressed (and briefly explained in the Appendix) that although $(-\nabla^{2})^{1/2}$ is a non-local operator,   it can be handled as a local one in the context of the harmonic extension  to an additional dimension. Exploiting this fact, one can prove  useful identities that allow to discuss relevant physical properties of the systems with dynamics governed by Lagrangians containing such operator.

A $z=2$  Lifshitz action
for   massive free  fermions was presented in eq.(\ref{DL})  and for massless fermions coupled to a $U(1)$ gauge field   in eq.(\ref{simple}). It should be pointed that ``squaring'' the resulting Dirac-Lifshitz equation (\ref{DLDL}) one gets the prototypical
equation for Lifshitz bosons (eq.(\ref{action}) with $z=2$)  in the same way as   the Klein-Gordon equation can be obtained by ``squaring''
 the Dirac one.

The $z=2$ Dirac-Lifshitz action  that we have proposed for massless fermions is invariant under chiral transformations and we have constructed
 the associated Nother current which leads to classically conserved charges. The $j_5^0$ component coincides with the usual chiral charge density arising in the relativistic case but the spatial components $j^i_5$ differ. They have however a very simple form given by eq.(\ref{analog}) for free fermions and eq.(\ref{analogous}) for fermions coupled to an Abelian gauge field. This should be contrasted with the $z=3$ case in which $j_i^5$ turns to be much more complicated \cite{Bakas:2011nq}. In any case the  coupling of gauge fields to Lifshitz fermions   makes  the resulting chiral current  depend on the gauge field even in the $U(1)$ case, in contrast with what happens with the relativistic theory. It should be noticed that the same phenomenon takes place for the case of $z=3$ theories constructed from
 local operators \cite{Bakas:2011nq, Bakas}.

As it was to be expected, there is an anomaly   at the quantum level  which we have calculated in the two-dimensional case,
showing that it takes the same form as in the ordinary Dirac theory.  The fact that the anomaly is the same for relativistic and Lifshitz fermions  is in agreement with the results in refs.
\cite{Dhar:2009am}-\cite{Bakas:2011nq,Bakas} for the $z=3$ case in $3+1$ dimensions, and should be related to the connection
between the obstruction to the chiral current conservation law  and the topological charge of the gauge field configuration.

In view of the connection between the anomaly in two dimensional space-time and bosonization (see \cite{GMSS}
and references therein) one could expect to find the rules to connect the $z=2$ fermionic theory we propose with a $z=2$ two-dimensional bosonic theory, possibly with action (\ref{action}). We hope to come back to this problem in a future work.

We discussed in section 4  fermionic Lifshitz theories  with $z=2$ within the  stochastic quantization framework, a natural one recalling  that in  Parisi and Wu approach \cite{Parisi} one passes from a $d$ dimensional Euclidean action    to a $d+1$ effective actions which in general exhibit anisotropic scaling. Since one can introduce in the Langevin equation different kernels acting on the drift force, one can find different effective $d+1$ actions which nevertheless correspond to the same
$d$-dimensional quantum theory. Exploiting this fact we were able to construct several
$z=2$  fermionic Lifshitz theories. Moreover, through the well-known connection between the Fokker-Plank partition function and supersymmetry, we were able to find the supersymmetric extensions of the resulting  fermionic $z=2$
theories.

~

~

\noindent \underline{Acknowledgements}: We would like to thank Horacio Falomir, Eduardo Fradkin,  Gustavo Lozano, Mariel Santangelo and Luis Silvestre for helpful comments and suggestions. F.A.S is associated to CICBA and partially supported
by CONICET, ANPCyT and CICBA grants. H.M. thanks to CONICET for finantial support.

\section{Appendix: Square root operators}

In order to handle  the operator $(-\nabla ^{2})^{1/2}$ on  $\mathbb{R}^{n}$ we
follow ref.\ \cite{Caffa} where it is defined  in terms of the harmonic
extension problem to the upper half space $\mathbb{R}^{n}\times
\left( 0, \infty \right)$ of functions on $\mathbb{R}^{n}$. That is, given a function $f:\mathbb{R}^{n}\mathbb{%
\longrightarrow }\mathbb{R}$, one looks for an harmonic function $u:\mathbb{R}^{n}\times
\left( 0, \infty \right)\mathbb{\longrightarrow }\mathbb{R}$ such that its restriction to $\mathbb{R}^{n}$ coincides with  $f$. Thus, the problem reduces to solve the
Dirichlet problem%
\begin{equation}
\left\{
\begin{array}{c}
u\left( x,0\right) =f\left( x\right)  \\
\\
\nabla _{n+1}^{2}u\left( x,y\right) =0%
\end{array}%
\right.   \label{1}
\end{equation}%
where $\nabla _{n+1}^{2}$ is the Laplacian operator in $\mathbb{R}^{n+1}$, $%
x\in \mathbb{R}^{n}$ and $y\in \mathbb{R}$. This is a well studied problem: for a smooth function on $C_0^ \infty (\mathbb{R}^{n})$ , there is a unique harmonic extension $u\in C^ \infty (\mathbb{R}^{n}\times
\left( 0, \infty \right))$.

One then introduces an operator $T$ defined on functions $f:\mathbb{R}^{n}%
\mathbb{\longrightarrow }\mathbb{R}$ having an harmonic extension $u:$ $%
\mathbb{R}^{n}\times
\left( 0, \infty \right)\mathbb{\longrightarrow }\mathbb{R}$,
such that%
\begin{equation}
\left( Tf\right) \left( x\right) =-\left. \frac{\partial u\left( x,y\right)
}{\partial y}\right\vert _{y=0}  \label{2}
\end{equation}%
Thus, since $\left( Tf\right) \left( x\right) $ has also an harmonic
extension to $\mathbb{R}^{n}\times
\left( 0, \infty \right))$, namely $u_{y}\left(
x,y\right) $ provided $u\left( x,y\right) $ is the harmonic extension of $f$%
, the sucessive applications of $T$ gives%
\begin{equation*}
\left( T\left( Tf\right) \right) \left( x\right) =\left. \frac{\partial
^{2}u\left( x,y\right) }{\partial y^{2}}\right\vert _{y=0}=-\left. \nabla
_{n}^{2}u\left( x,y\right) \right\vert _{y=0}=-\nabla _{n}^{2}f\left(
x\right)
\end{equation*}%
that is equivalent to write%
\begin{equation*}
T^{2}f\left( x\right) =\left( -\nabla _{n}^{2}\right) f\left( x\right)
\end{equation*}%
Then one can  identify
\begin{equation}
T=\left( -\nabla _{n}^{2}\right) ^{1/2}  \label{3}
\end{equation}

Let us consider the case $n=1$, so $f:\mathbb{%
R\longrightarrow }\mathbb{R}$ and we look for harmonic extensions to $%
\mathbb{R}\times
\left( 0, \infty \right)$. For instance, taking $f\left(
x\right) =\cos kx$, it is easy to see that $u\left( x,y\right)
=e^{-\left\vert k\right\vert y}\cos kx$, $y\in \left( 0, \infty \right)$, is
a  {bounded} harmonic extension of $f\left( x\right) $,
\begin{equation*}
\left( \frac{\partial ^{2}}{\partial x^{2}}+\frac{\partial ^{2}}{\partial
y^{2}}\right) u\left( x,y\right) =-k^{2}e^{-\left\vert k\right\vert y}\cos
kx+\left\vert k\right\vert ^{2}e^{-\left\vert k\right\vert y}\cos kx=0
\end{equation*}%
Hence, we may evaluate the action of the operator $\left( -\nabla
_{1}^{2}\right) ^{1/2}=\left( -\partial _{1}^{2}\right) ^{1/2}$ on $f$%
\begin{equation*}
\left( -\partial _{1}^{2}\right) ^{1/2}\cos kx=T\left( \cos kx\right)
=-\left. u_{y}\left( x,y\right) \right\vert _{y=0}=\left\vert k\right\vert
\cos kx
\end{equation*}%
The same arguments hold if one takes $f =  \sin kx$ since  the function $v\left(
x,y\right) =e^{-\left\vert k\right\vert y}\sin kx$ is a  {bounded}
harmonic extension on $\mathbb{R}\times \left( 0, \infty \right)$ of $\sin kx$%
\begin{equation*}
\left( \frac{\partial ^{2}}{\partial x^{2}}+\frac{\partial ^{2}}{\partial
y^{2}}\right) v\left( x,y\right) =-k^{2}e^{-\left\vert k\right\vert y}\sin
kx+\left\vert k\right\vert ^{2}e^{-\left\vert k\right\vert y}\sin kx=0
\end{equation*}%
and%
\begin{equation*}
\left( -\partial _{1}^{2}\right) ^{1/2}\sin kx=T\left( \sin kx\right)
=-\left. v_{y}\left( x,y\right) \right\vert _{y=0}=\left\vert k\right\vert
\sin kx
\end{equation*}
We can then write the action of $\left( -\partial
_{1}^{2}\right) ^{1/2}$ on $e^{ikx}$ as%
\begin{equation*}
\left( -\partial _{1}^{2}\right) ^{1/2}e^{ikx}=\left( -\partial
_{1}^{2}\right) ^{1/2}\cos kx+i\left( -\partial _{1}^{2}\right) ^{1/2}\sin kx
\end{equation*}%
or
\begin{equation}
 \left( -\partial _{1}^{2}\right) ^{1/2}e^{ikx}=\left\vert
k\right\vert e^{ikx}   \label{exp}
\end{equation}
this proving eq.(16).

~

\noindent{\bf Remark 1:}   Suppose that the function $f\left( x\right) $
admits an harmonic extension $u\left( x,y\right) $, so $\left( Tf\right) \left(
x\right) =-u_{y}\left( x,0\right) $. Then,  functions $\partial _{i}f\left( x\right) $, $i=1,...,n$, have harmonic extensions
$\partial _{i}u\left( x,y\right) $ and
\begin{eqnarray*}
\left( T\left( \partial _{i}f\right) \right) \left( x\right)  &=&-\left.
\partial _{y}\left( \partial _{i}u\left( x,y\right) \right) \right\vert
_{y=0} \\
&=&-\partial _{i}\left. u_{y}\left( x,y\right) \right\vert _{y=0} \\
&=&\partial _{i}\left( Tf\right) \left( x\right)
\end{eqnarray*}%
provided that  $u\in C^{2}\left( \mathbb{R}\times
\left( 0, \infty \right))\right) $. From this, we conclude that
\begin{equation}
\left( -\nabla _{n}^{2}\right) ^{1/2}\partial _{i}=\partial
_{i}\left( -\nabla _{n}^{2}\right) ^{1/2}   \label{4}
\end{equation}

\noindent{\bf Remark 2:} We now address the selfadjointness of the operator $T$, following reference \cite{Tan}. Let $%
u\left( x,y\right) $ and $v\left( x,y\right) $ harmonic extensions to $%
\mathcal{C}=\mathbb{R}^{n}\times
\left( 0, \infty \right))$ of $g\left( x\right) $ and
$f\left( x\right) $, respectively. Also, assume that both the harmonic
extensions vanish for $\left\vert x\right\vert ,\left\vert y\right\vert
\longrightarrow \infty $, then%
\begin{eqnarray}
\int_{\mathcal{C}} \nabla _{n+1}u\left( x,y\right)  \cdot  \! \nabla
_{n+1}v\left( x,y\right) dxdy \!\!\! &=&
\int_{\mathcal{C}}\!\!\nabla _{n+1}\! \cdot\!  \left( u\left( x,y\right) \nabla
_{n+1}v\left( x,y\right) \right) dxdy  \notag \\
&=&\int_{\partial \mathcal{C}}u\left( x,y\right) \nabla _{n+1}v\left(
x,y\right) ~dxdy  \notag \\
&=&-\int_{\mathbb{R}^{n}}\left. u\left( x,y\right) \frac{\partial }{\partial
y}v\left( x,y\right) \right\vert _{y=0} dx  \label{integral over C}
\end{eqnarray}%
Here, $\partial \mathcal{C}$  indicates the border of $\mathcal{C}$.  To obtain this
identity
we have written the  integral of the $\left(
n+1\right) $-dimensional divergence $\left( \ref{integral over C}\right) $
as  a  {surface integral } over the component of the vector
field normal to the {surface }$\partial
\mathcal{C}$ which is taken as  a finite $\left( n+1\right) $-dimensional cube with a face
at $y=0$. Then we have made  the $2n+1$ faces other than  $y=0$ to go to
infinity, where  functions $u$
and $v$ vanish.

 Hence, for the harmonic extensions $u$,$v$
to $%
\mathcal{C}=\mathbb{R}^{n}\times
\left( 0, \infty \right)$  of $g,f $, respectively, we have%
\begin{equation*}
\int_{\mathcal{C}}u\left( x,y\right) \nabla _{n+1}^{2}v\left( x,y\right)
~dxdy-\int_{\mathcal{C}}v\left( x,y\right) \nabla _{n+1}^{2}u\left(
x,y\right) ~dxdy=0
\end{equation*}%
or, using (\ref{integral over C}),
\begin{equation*}
\int_{\mathbb{R}^{n}}\left. \left( u\left( x,y\right) \frac{\partial }{%
\partial y}v\left( x,y\right) -v\left( x,y\right) \frac{\partial }{\partial y%
}u\left( x,y\right) \right) \right\vert _{y=0}~dx=0
\end{equation*}%
In terms of $g\left( x\right) $ and $f\left( x\right) $ this identity  reads  %
\begin{equation*}
\int_{\mathbb{R}^{n}}\left( g\left( x\right) \frac{\partial f\left( x\right)
}{\partial y}-\frac{\partial g\left( x\right) }{\partial y}f\left( x\right)
\right) ~dx=0
\end{equation*}%
or, equivalently%
\begin{equation}
\int_{\mathbb{R}^{n}}g\left( x\right) Tf\left( x\right) dx=\int_{%
\mathbb{R}^{n}}f\left( x\right) Tg\left( x\right) dx
\label{self adjointness}
\end{equation}%

Concerning the case in which    fermions are coupled to the $U(1)$ gauge field, we introduce  the operator ${\cal T}[A]$ defined as
\be
{\cal T}[A] = (-D_i[A]D_i[A])^{1/2}
\ee
with the covariant derivative given by
\be
D_i = i\partial_i + e A_i
\ee
In the case of $ d=1$ space dimensions  this simplifies to
\be
{\cal T}[A] = (-D_1[A]D_1[A])^{1/2}
\ee

The covariant derivative $D_1[A]$  can be written as
\be
D_1[A] = \exp\left(
ie\int^x A_1 dy
\right)
\left(i\frac{d}{dx}\right)
\exp\left(
-ie\int^x A_1 dy
\right)
\label{6}
\ee
which in terms of a Wilson line
\be
W[A_1] = \exp\left(
-ie\int^x A_1 dy
\right)
\ee
takes the compact form
\be
D_1[A] = W^{-1}[A_1] i\frac{d}{dx} W[A_1]
\ee
One then has
\be
\left(D_1[A]\right)^2 = W^{-1}[A_1] \left(i\frac{d}{dx}\right)^2 W[A_1] =  W^{-1}[A_1] T^2 W[A_1]
\ee
or
\be
\left(D_1[A]\right)^2= W^{-1}[A_1] T W[A_1]W^{-1}[A_1] T W[A_1]
\ee
so that
\be
{\cal T}[A_1] = W^{-1}[A_1] T W[A_1]
\ee
showing that
${\cal T}[A_1] $ is Hermitian.


\newpage
\begin{thebibliography}{99}
\bibitem{H}
  R.~M.~Hornreich, M.~Luban and S.~Shtrikman,
  Phys.\ Rev.\ Lett.\  {\bf 35},  1678  (1975).
  \bibitem{Gri} G.~Grinstein, Phys.\ Rev.\  {\bf B23}, 4615 (1981) .
  \bibitem{AF}
  E.~Ardonne, P.~Fendley and E.~Fradkin,
  Annals Phys.\  {\bf 310},  493 (2004)
  [cond-mat/0311466].
  \bibitem{Frad} E.~Fradkin,
   J.\ of Phys.\ {\bf A42}, 504011 (2009).
  \bibitem{Horava0}
  P.~Ho\v{r}ava,
  Phys.\ Lett.\ B {\bf 694},172 (2010)
  [arXiv:0811.2217 [hep-th]].
  \bibitem{Horava}
P. Ho\v{r}ava, Phys.\ Rev.\ D79, 084008 (2009) [arXiv:0901.3775];
JHEP 03, 020 (2009) [arXiv:0812.4287];
\bibitem{Horava2}
  P.~Ho\v{r}ava, Phys.\ Lett.\  B {\bf 694},  172  (2010) [arXiv:0811.2217];
Phys.\ Rev.\ Lett.\ 102, 161301  (2009) [arXiv:0902.3657].
\bibitem{Alex} J.~Alexandre,
  Int.\ J.\ Mod.\ Phys.\ A {\bf 26}, 4523  (2011)
  [arXiv:1109.5629 [hep-ph]].
  \bibitem{Kachru}
  S.~Kachru, X.~Liu and M.~Mulligan,
  Phys.\ Rev.\ D {\bf 78}, 106005  (2008)
  [arXiv:0808.1725 [hep-th]].
\bibitem{fk} Kai Sun, Hong Yao,  Eduardo Fradkin and Steven A. Kivelson,
Phys.\ Rev.\ Lett.\ 103,  046811 (2009).
  \bibitem{Anselmi1}
  D.~Anselmi and M.~Halat,
  Phys.\ Rev.\ D {\bf 76}, 125011 (2007)
  [arXiv:0707.2480 [hep-th]].
\bibitem{Dhar:2009dx}
  A.~Dhar, G.~Mandal and S.~R.~Wadia,
  ``Asymptotically free four-fermi theory in 4 dimensions at the z=3 Lifshitz-like fixed point,''
  Phys.\ Rev.\ D {\bf 80}, 105018 (2009)
  [arXiv:0905.2928 [hep-th]].
  \bibitem{Dhar:2009am}
  A.~Dhar, G.~Mandal and P.~Nag,
  Phys.\ Rev.\ D {\bf 81} (2010) 085005
  [arXiv:0911.5316 [hep-th]].
  \bibitem{Anselmi2}
  D.~Anselmi,
  Eur.\ Phys.\ J.\ C {\bf 65}, 523  (2010)
  [arXiv:0904.1849 [hep-ph]].
  \bibitem{Bakas:2011nq}
  I.~Bakas and D.~Lust,
  Fortsch.\ Phys.\  {\bf 59}, 937 (2011)
  [arXiv:1103.5693 [hep-th]].
    \bibitem{Bakas}  I.~Bakas,
  arXiv:1110.1332 [hep-th].
  \bibitem{AlexandreU}
  J.~Alexandre, J.~Brister and N.~Houston,
  arXiv:1204.2246 [hep-ph].
  \bibitem{Caffa} L.~Caffarelli and L.~Silvestre,
   Comm.\ in Part.\ Diff.\ Eqs.\ {\bf 32}, 1245   (2007)  [arXiv:math/0608640v2].
\bibitem{Seeley}
  R.~T.~Seeley,
  Proc.\ Symp.\ Pure Math.\  {\bf 10},  288 (1967).
    \bibitem{Laemmerzahl:1993xe}
  C.~Laemmerzahl,
  J.\ Math.\ Phys.\  {\bf 34},  3918 (1993).
    \bibitem{Giambiagi}
  J.~J.~Giambiagi,
  Nuovo Cim.\ A {\bf 104}, 1841 (1991).
\bibitem{Bollini}
  C.~G.~Bollini and J.~J.~Giambiagi,
  J.\ Math.\ Phys.\  {\bf 34}, 610 (1993).
  \bibitem{Shubin} M.~A.~Shubin,
  {"Pseudodifferential Operators and Spectral Theory"}, Springer, 1987.
  \bibitem{Tan} J.\ Tan,
  Calc.\ Var {\bf 42}, 21 (2011).
\bibitem{Fujikawa}
  K.~Fujikawa,
  Phys.\ Rev.\ Lett.\  {\bf 42}, 1195 (1979).
  \bibitem{Fujikawa2}
  K.~Fujikawa,
  Phys.\ Rev.\ D {\bf 21}, 2848  (1980)
   [Erratum-ibid.\ D {\bf 22}, 1499 (1980)].
   \bibitem{GMSS}
  R.~E.~Gamboa Saravi, M.~A.~Muschietti, F.~A.~Schaposnik and J.~E.~Solomin,
  Annals Phys.\  {\bf 157}, 360  (1984) .
  \bibitem{Dijk}
  R.~Dijkgraaf, D.~Orlando and S.~Reffert,
  Nucl.\ Phys.\ B {\bf 824}, 365  (2010)
  [arXiv:0903.0732 [hep-th]].
 \bibitem{Orl}
  D.~Orlando and S.~Reffert,
  Phys.\ Lett.\ B {\bf 683}, 62  (2010)
  [arXiv:0908.4429 [hep-th]].
  \bibitem{Parisi}
  G.~Parisi and Y.~-s.~Wu,
  Sci.\ Sin.\  {\bf 24},  483 (1981).
  \bibitem{Fukai}
  T.~Fukai, H.~Nakazato, I.~Ohba, K.~Okano and Y.~Yamanaka,
  Prog.\ Theor.\ Phys.\  {\bf 69}, 1600  (1983).
  \bibitem{DT}P.~H.~Damgaard and K.~Tsokos,
  Nucl.\ Phys.\ B {\bf 235},  75  (1983).
  \bibitem{vanN} U.~Lindstr\"om, M.~Rocek, W.~Siegel, P.~van Nieuwenhuizen, and A. E. van de Ven,
J.\ Math.\ Phys.\ {\bf 31}, 1761  (1990).
\bibitem{ParSour}   G.~Parisi and N.~Sourlas,
  Nucl.\ Phys.\ B {\bf 206},  321 (1982).
\bibitem{Cha} S.~Chaturvedi, A.~K.~Kapoor and V.~Srinivasan,
  Phys.\ Lett.\ B {\bf 140}, 56 (1984).
 \end{thebibliography}
\end{document}